\documentclass{iopart}

\usepackage{graphicx}

\begin{document}

\title{Switching Dynamics in Reaction Networks Induced by Molecular Discreteness}
\author{Yuichi Togashi$^1$ and Kunihiko Kaneko$^{2,3}$}
\address{$^1$ Department of Physical Chemistry, Fritz Haber Institute of the Max Planck Society,
Faradayweg 4-6, 14195 Berlin, Germany}
\address{$^2$ Department of Basic Science, School of Arts and Sciences, The University of Tokyo,
Komaba, Meguro, Tokyo 153-8902, Japan}
\address{$^3$ ERATO Complex Systems Biology Project, JST, Komaba, Meguro, Tokyo 153-8902, Japan}
\ead{togashi@fhi-berlin.mpg.de}

\date{\today}

\begin{abstract}

To study the fluctuations and dynamics in chemical reaction processes, stochastic differential equations based on the rate equation involving chemical concentrations are often adopted. When the number of molecules is very small, however, the discreteness in the number of molecules cannot be neglected since the number of molecules must be an integer. This discreteness can be important in biochemical reactions, where the total number of molecules is not significantly larger than the number of chemical species. To elucidate the effects of such discreteness, we study autocatalytic reaction systems comprising several chemical species through stochastic particle simulations. The generation of novel states is observed; it is caused by the extinction of some molecular species due to the discreteness in their number. We demonstrate that the reaction dynamics are switched by a single molecule, which leads to the reconstruction of the acting network structure. We also show the strong dependence of the chemical concentrations on the system size, which is caused by transitions to discreteness-induced novel states.

\end{abstract}

\submitto{\JPCM}

\maketitle

\section{Introduction}

In nature, there exist various systems that involve
chemical reactions.
Some systems are on a geographical scale while others are on a nanoscale,
in particular, the biochemical reactions in a cell.
To study the dynamics of reaction systems, we often adopt rate equations
in order to observe the change in the chemical concentrations.
In rate equations, we consider the concentrations to be continuous variables
and the rate of each reaction as a function of the concentrations.
In fact, in macroscopic systems, there are a large number of molecules;
therefore, continuous representations are usually applicable. 

When the concentration of a certain chemical is low,
fluctuations in the reactions or flows cannot be negligible.
They are usually treated by employing stochastic differential equations,
in which the noise is used as a continuum description of the fluctuations
\cite{Nicolis,Kampen}.
The employment of stochastic differential equations has led to some important discoveries
such as noise-induced transitions \cite{NIP}, noise-induced order \cite{NIO},
and stochastic resonance \cite{SR}.

In stochastic differential equations, the quantities of chemicals are still
regarded as continuous variables.
At a microscopic level, however, we need to seriously consider the fact that
the number of molecules is an integer ($0$, $1$, $2$, $\cdots$)
that changes in a discrete manner.
Fluctuations originate from the discrete stochastic changes;
thus, continuum descriptions of fluctuations are not always appropriate.

Biological cells appear to provide a good example for such discreteness
in molecule numbers. The size of the cells is of the order of microns,
in which nanoscale ``quantum'' effects can be ignored.
However, in these cells, some chemicals act
at extremely low concentrations of the order of pM or nM.
Assuming that the typical volume of a cell ranges from $1$ to
$10^{3}$ $\mu m^{3}$, the concentration of
one molecule in the cell volume corresponds to $1.7$ pM to $1.7$ nM.
It is possible that the molecule numbers of some chemicals
in a cell are of the order of $1$ or sometimes even $0$.

If such chemicals play only a minor role in a cell,
we can safely ignore these chemicals to study intracellular chemical processes.
However, this is not always the case.
In biological systems, chemical species with a small number of
molecules may critically affect the behavior of the entire system.
As an extreme example, there exist only one or a few copies of genetic molecules
such as DNA, which are important to characterize the behavior of each cell.
Further, some experiments show that doses of particular chemicals
at concentrations of the order of pM or fM may alter cell behavior
(e.g., \cite{TGF1,IL1B}).
Biological systems also include positive-feedback mechanisms
such as autocatalytic reactions,
which may amplify single molecular changes to a macroscopic level.
It is only recently that the stochastic effect due to small molecule numbers
in cells has been noticed
both theoretically \cite{McAdams,Arkin1} and experimentally \cite{Elowitz2002}.

In this paper, we focus on the possible effects of molecular discreteness.
Through stochastic simulations,
we showed that the discreteness can induce transitions to novel states in
autocatalytic systems \cite{YTKK2001},
which may affect macroscopic chemical concentrations \cite{YTKK2003}.
In the first part of this paper,
we briefly review these studies and explain other aspects of such effects.
See also \cite{Bettelheim,Marion,CaOscillator} for recent advances
in the present topic by analytic methods and numerical simulations and
\cite{Gibson,GFRD} for simulation methods concerned.

In some cases, the discreteness in the molecule numbers
may cause switches between two or more states with distinct
concentrations and dynamical behaviors.
Further, even though the concentration of chemicals is sufficiently high
for one state, the concentration could be low in another state,
in which a chemical with a very low concentration could work as a stochastic switch.
In the second part of this paper, we discuss how molecular discreteness leads to switch
among states with distinct dynamical behaviors in an autocatalytic chemical
reaction network system.
This spontaneous switching is characterized as an alteration
(i.e., disconnection and reconnection) of the acting reaction paths.

\section{Discreteness-induced transitions and alteration of concentrations}

\begin{figure}
\begin{center}
\includegraphics[width=80mm]{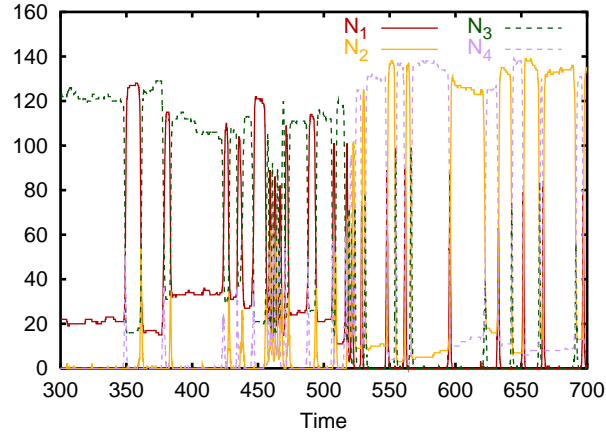}
\end{center}
\caption{Time series of $N_{i}$ for $V=32$, $D_{i}=1/256$, and $r_{i}=s_{i}=1$.
In this case, $N_{i}$ can reach $0$,
and the switching states appear.
In the 1-3 rich state, the system successively switches
between the $N_{1} > N_{3}$ and $N_{1} < N_{3}$ states.
The interval of switching is considerably longer than
the period of continuous vibration ($\approx \pi$).
At around $t=520$, a transition occurs from the 1-3 rich state to the 2-4 rich state.}
\label{fig:A-032}
\end{figure}

\begin{figure}
\begin{center}
\includegraphics[width=64mm]{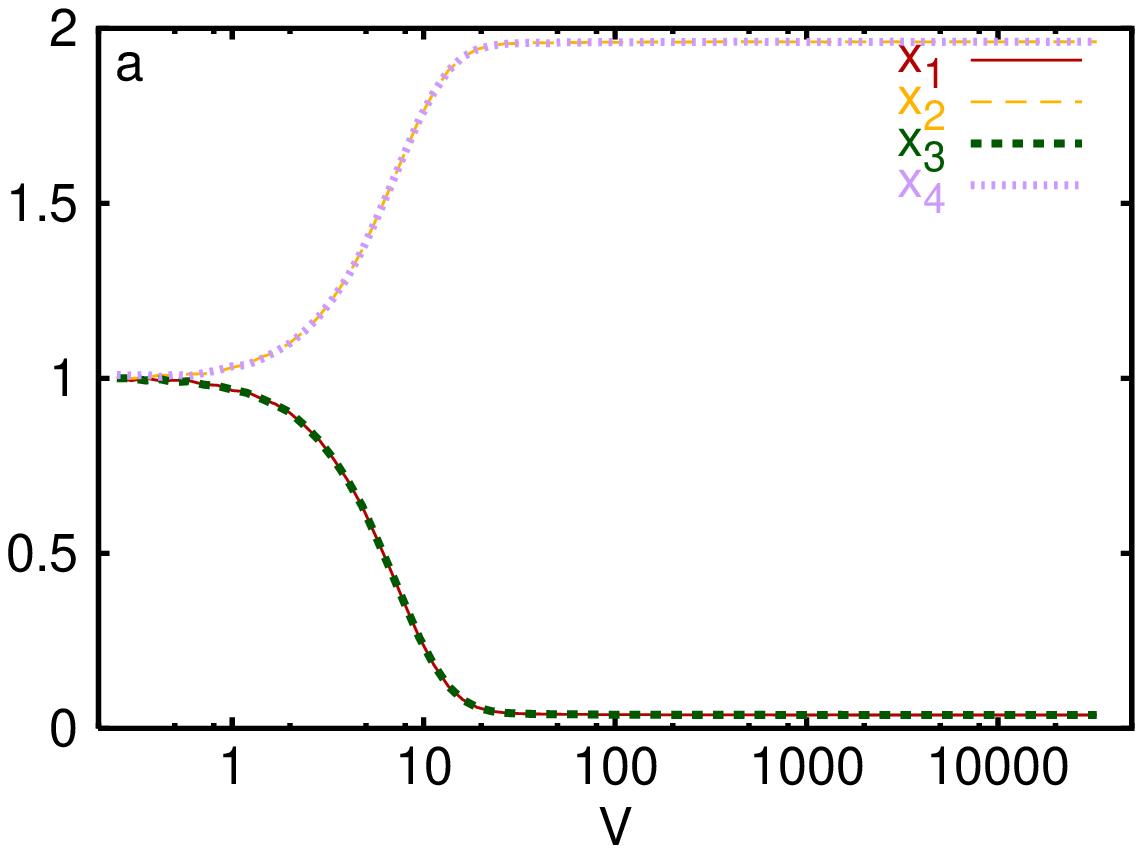}
\includegraphics[width=64mm]{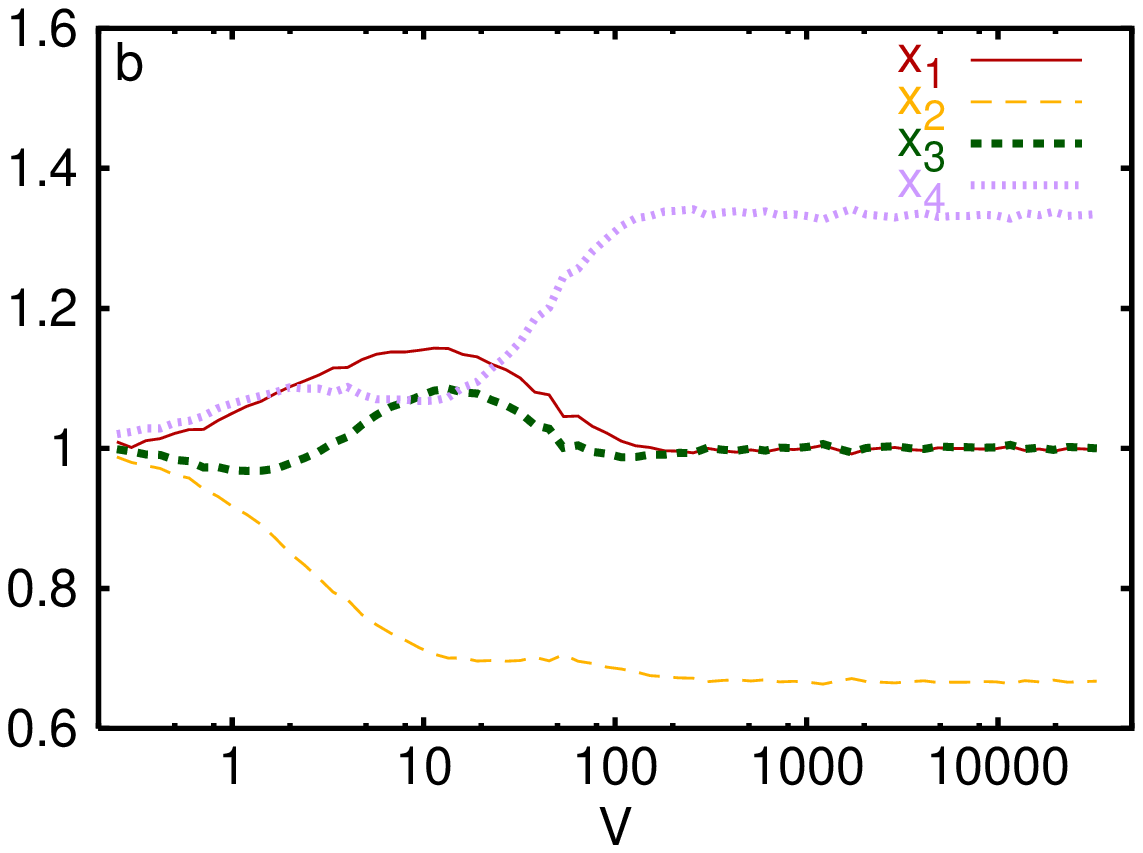}
\end{center}
\caption{The average concentrations $\bar{x_{i}}$ in the autocatalytic loop system
for $\forall i : s_{i}=1$ and $D_{i}=1/128$ with nonequivalent reaction constants.
For small $V$, the flow of molecules dominates the system.
Thus, $\bar{x_{i}} \approx 1$, which simply reflects $s_{i}=1$;
this does not depend on how the continuum limit is imbalanced by the reactions.
(a) $r_{1}=r_{3}=1$ and $r_{2}=r_{4}=0.9$.
(b) $r_{1}=r_{2}=2$ and $r_{3}=r_{4}=1$.}
\label{fig:tr2-average-r}
\end{figure}

We have previously reported that the discrete nature of molecules may induce
transitions to novel states, which are not reproduced by the continuous
descriptions of the dynamics (stochastic differential equations) \cite{YTKK2001, YTKK2003}.
Here, we briefly review that result by including some novel results.

We consider a simple autocatalytic network (loop) with four
chemicals $X_{i}$ ($i=1,\cdots,4$).
We assume the reactions
$X_{i} + X_{i+1} \rightarrow 2X_{i+1}$ (with $X_{5} \equiv X_{1}$)
between these chemicals.
All the reactions are irreversible.

We assume that the reactor is a well-stirred container with a volume $V$.
The set of $N_{i}$, the number of $X_{i}$ molecules, determines the state
of the system.
The container is in contact with a chemical reservoir in which
the concentration of $X_{i}$ is fixed at $s_{i}$.
The flow rate of $X_{i}$ between the container and the reservoir is $D_{i}$,
which corresponds to the probability of the flow-out of a molecule
per unit time \footnote{$D_{i}$ is the diffusion rate across the surface
of the container.
Here, we choose a flow proportional to $V$ in order to
obtain a well-defined continuum limit.}.

We can consider the continuum limit as $V \rightarrow \infty$.
In this limit, the change in $x_{i}$,
the chemical concentration of $X_{i}$ in the container,
obeys the following rate equation:
\begin{equation}
\frac{dx_{i}}{dt} = r_{i-1}x_{i-1}x_{i} - r_{i}x_{i}x_{i+1} + D_{i}(s_{i} - x_{i}),
\label{eqn:tr1-rate}
\end{equation}
where $r_{i}$ is the rate constant of
the reaction $X_{i} + X_{i+1} \rightarrow 2X_{i+1}$, and $X_{0} \equiv X_{k}$.

In \cite{YTKK2001}, we considered a case with four equivalent chemical species,
given as $r_{i}=r$, $D_{i}=D$, and $s_{i}=s$ for all $i$ ($r$, $D$, $s > 0$), $k=4$.
In the continuum limit, the dynamics is represented by the rate equation,
which has only one attractor: a stable fixed point $x_{i} = s$ for all $i$.
Around the fixed point, $x_{i}$ vibrates
with the frequency $\omega_{p} \equiv rs / \pi$.
If the number of molecules is finite but fairly large,
we can estimate the dynamical behavior of the system
using the Langevin equation, which is obtained by adding a noise term to the rate equation.
Each concentration $x_{i}$ fluctuates and vibrates around the fixed point.
An increase in the noise (corresponding to a decrease in the number of molecules)
merely amplifies the fluctuation.

However, as we have shown in \cite{YTKK2001},
when the number of molecules is small, novel states
that do not exist in the continuum limit are observed.
Two chemicals are dominant and the other two are mostly extinct ($N_{i}=0$).
Figure \ref{fig:A-032} shows the time series of $N_{i}$ in such a case.
At $t < 520$, $N_{1}$ and $N_{3}$ dominate the system and
$N_{2} = N_{4} = 0$ for the most part (the 1-3 rich state).
Once the system reaches $N_{2} = N_{4} = 0$, all the reactions stop.
The system remains at $N_{2} = N_{4} = 0$ for a long time as compared with
the ordinary time scale of the reactions ($\sim 1/rs$).
Conversely, at $t > 520$, $N_{2}$ and $N_{4}$ are large and
usually $N_{1} = N_{3} = 0$ (the 2-4 rich state).
In the 1-3 or 2-4 rich states, the system alternately switches
between either $N_{1} > N_{3}$ and $N_{1} < N_{3}$ or
$N_{2} > N_{4}$ and $N_{2} < N_{4}$.
We name these states the ``switching states.''

The appearance of discreteness-induced novel states is described
as a phase transition with a decrease in the system size (or flow rate),
where the histogram of $(N_{1} + N_{3}) - (N_{2} + N_{4})$ exhibits
a change from single-peaked to double-peaked distribution.

In this example, although the state at each instant exhibits a clear transition,
the average concentrations of the chemicals are not altered since the system
resides equally
in the 1-3 rich and 2-4 rich states over a long time span.
On the other hand, we found examples in \cite{YTKK2003},
where the long-term average concentration of the molecule species is altered
by the discreteness as well.
A simple example is provided by considering the same reaction model
as that depicted by eq. (\ref{eqn:tr1-rate}), but by considering the case where
the parameters $D_{i}$, $s_{i}$, or $r_{i}$ are species dependent.
Note that the rate equation (\ref{eqn:tr1-rate}) obtained
in the continuum limit does not contain the volume $V$;
hence, the average concentrations should be independent of $V$.
Here, we seek the possibility of the change in the average
concentrations depending on the decrease in the system size $V$ by
taking advantage of the switching states. 

Recall that for the transitions to the switching states
to occur in \cite{YTKK2001}, it was necessary
for the interval of the inflow to be greater than
the time scale of the reactions.
In the present case, the inflow interval of $X_{i}$ is $\sim 1/D_{i}s_{i}V$,
and the time scale of the reaction $X_{i} + X_{i+1} \rightarrow 2X_{i+1}$ in order to
consume $X_{i}$ is $\sim 1/r_{i}x_{i+1}$.
If the conditions of all the chemicals are equivalent,
the discreteness of all the chemicals takes effect equally
and the 1-3 and 2-4 rich states coordinately appear at $V \approx r / D$.

Now, since the parameters are species dependent,
the effect of the discreteness may be different for each species.
For example, by assuming that $D_{1}s_{1} < D_{2}s_{2}$,
the inflow interval of $X_{1}$ is greater than that of $X_{2}$.
Thus, the discreteness in the inflow of $X_{1}$ can be significant for larger $V$.

In the previous paper \cite{YTKK2003}, we studied the case in which only
the external concentrations (chemical inflows) $s_{i}$ were dependent on the species.
Based on the degree of $s_{i}V$, the discreteness-induced transition occurs
successively with the decrease in $V$,
and the average concentrations of the chemicals take distinct values
from those of the continuum limit case.
Similarly, we can study the effect of the discreteness
when each of the chemical reaction rates $r_{i}$ is species dependant.
In fact, the dependence of $\bar{x_{i}}$ on $V$ in this case is different from
the previous study in which only $s_{i}$ was species dependent.

For example, we consider the case that $r_{1} = r_{3} > r_{2} = r_{4}$ and
$\forall i: s_{i}=1$.
Figure \ref{fig:tr2-average-r} shows the dependence of
$\bar{x_{i}}$ on $V$.
Recall that the concentrations should not depend on $V$ as long as
the continuum representations hold (eq. (\ref{eqn:tr1-rate}) does not contain $V$).
Here, in the continuum limit or in the case of large $V$,
$\bar{x_{2}} = \bar{x_{4}} > \bar{x_{1}} = \bar{x_{3}}$,
as shown in Fig. \ref{fig:tr2-average-r} (a).
In contrast, when $V$ is small, $\bar{x_{i}} \approx 1$.
If $V$ is very small, so that the total number of molecules is mostly $0$ or $1$,
the reactions rarely occur and the flow of chemicals dominates the system.
Thus, $\bar{x_{i}} \approx s_{i}$.

If both the reactions and the flows are species dependent,
we can simply expect the behavior to be a combination of the above mentioned cases.
Even this simple system can exhibit a multi-step change in the concentrations
along with the change in $V$. Furthermore, the present behavior is not limited
to the simple autocatalytic reaction loop.
In fact, we observe this type of change
in randomly connected reaction networks.
For a large reaction network with multiple time scales of reactions and flows,
the discreteness effect may bring about behaviors that are more complicated,
although our discussion is generally applicable to such cases
if the time scales are appropriately defined.

As seen above, the discreteness of molecules can alter the average concentrations.
When the rates of inflow and/or the reaction are species dependent,
the transitions between the discreteness-induced states are imbalanced.
This may drastically alter the average concentrations
from those of the continuum limit case.

Note that although the concentrations are altered in both cases,
their dependence on $V$ is different.
If the system is extremely small ($V \sim 1$),
the frequency of the reaction event, in comparison to the diffusion, is low.
The reaction is limited by the inflows,
and therefore, the system is dominated by diffusion.
The average concentrations $\bar{x_{i}}$ depend on $V$, but the dependence
is quite different from the case with uniform reaction rates ($r_{i}$) and imbalanced
inflows ($s_{i}$), which were previously reported by us.

\section{Discreteness-induced switching of catalytic reaction networks}

\begin{figure}
\begin{center}
\includegraphics[width=128mm]{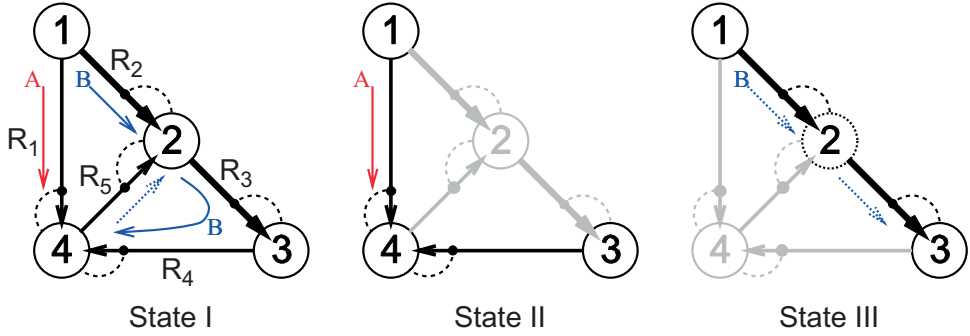}
\end{center}
\caption{Model catalytic network.
There are two reaction paths---indicated by arrows A and B---from
the chemical $X_{1}$ (substrate).
(I) If all the chemical species exist, all the reactions may occur.
The system exhibits damped oscillations.
(II,III) If the system lacks one or more chemicals,
some of the reactions cannot proceed.
The portion of the reaction path beyond the stalled reaction is disconnected;
consequently, the actual topology of the network may change.
}
\label{fig:net2}
\end{figure}

\begin{figure}
\begin{center}
\includegraphics[width=64mm]{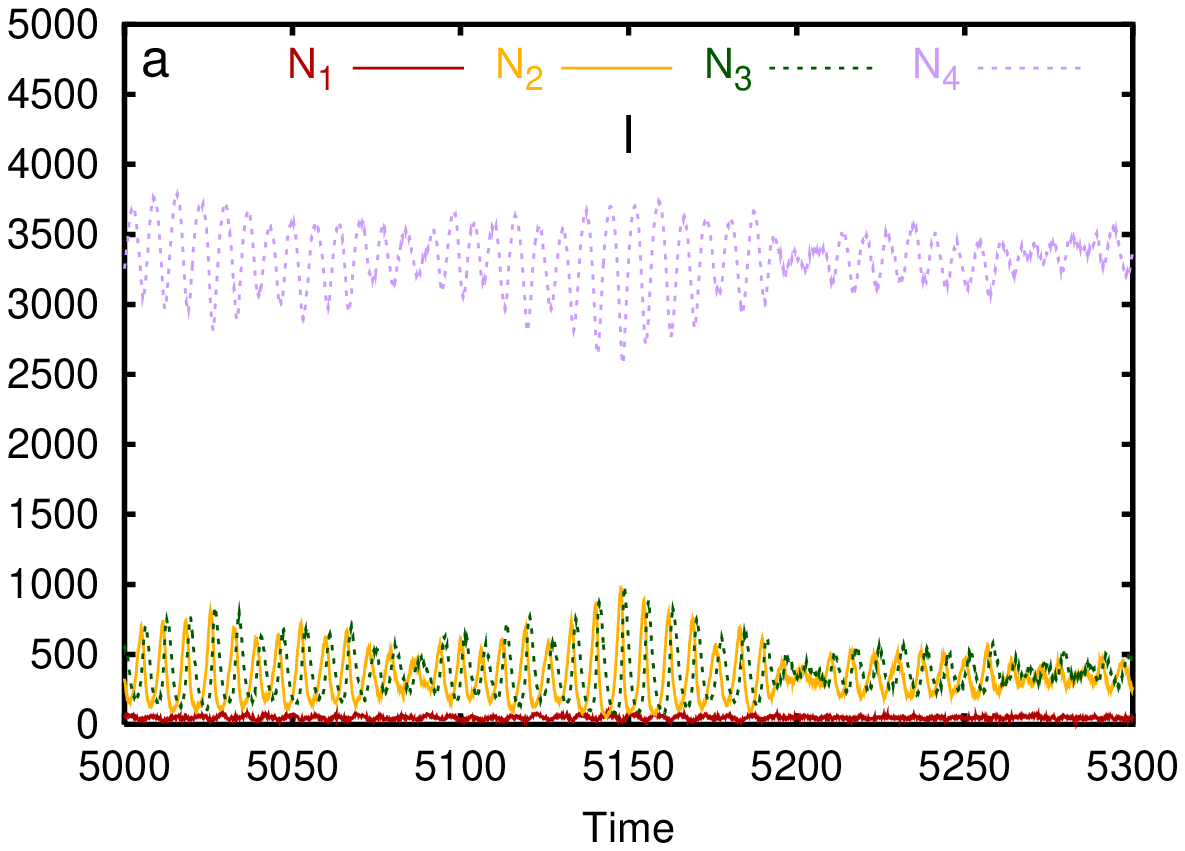}
\includegraphics[width=64mm]{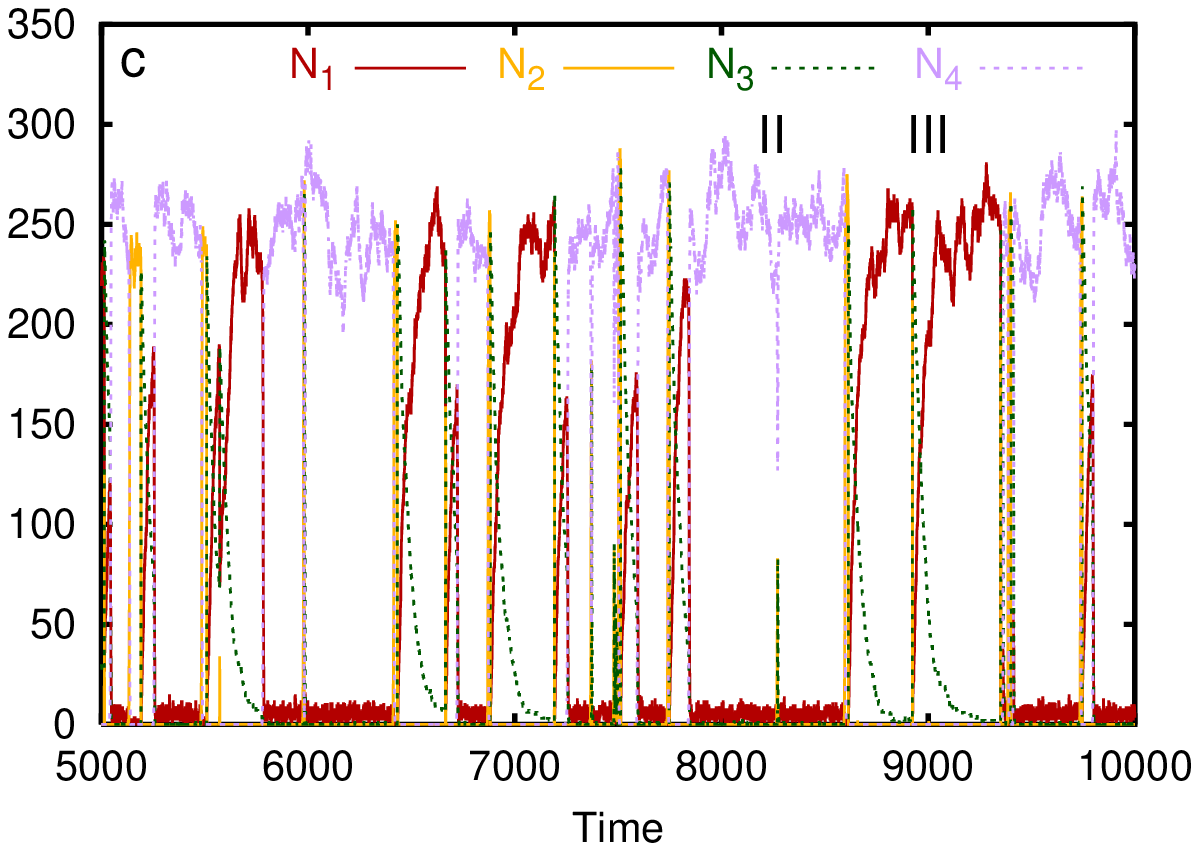}\\
\includegraphics[width=90mm]{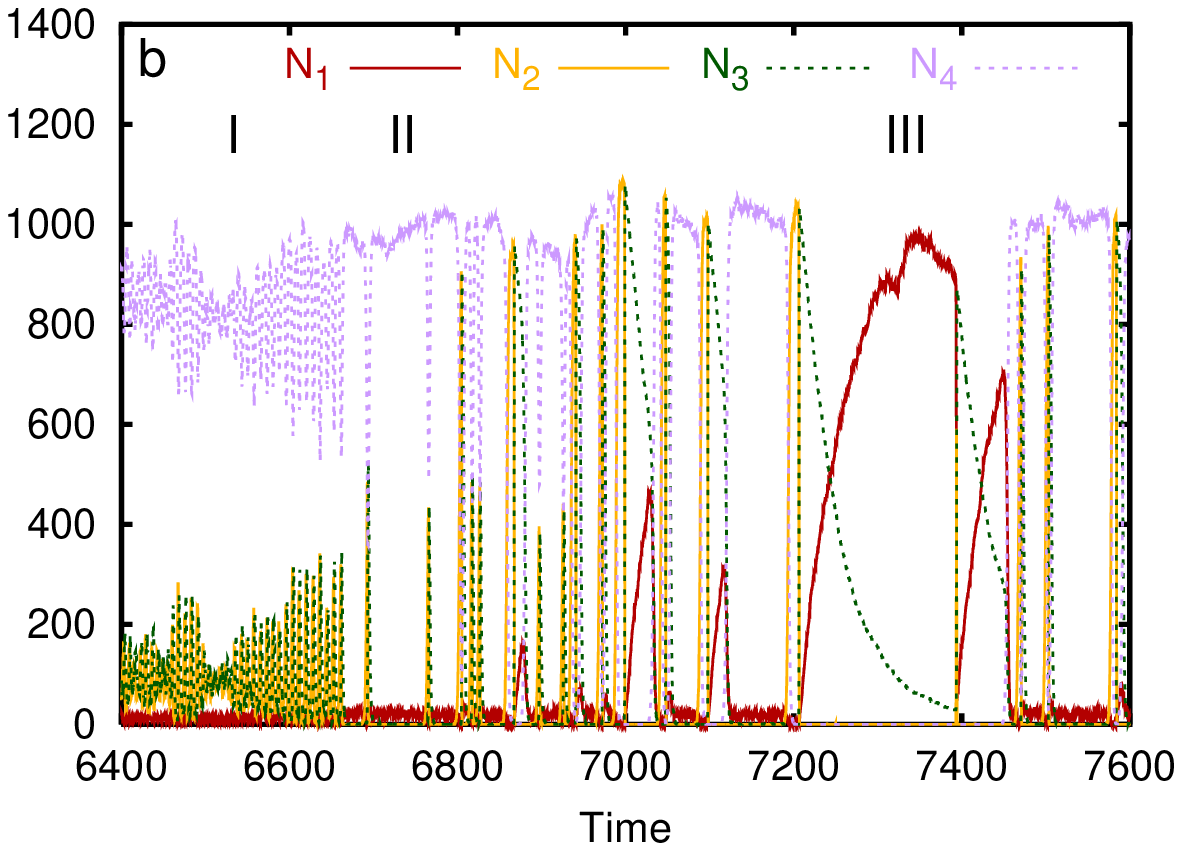}
\end{center}
\caption{Time series of $N_{i}$, the number of molecules.
(a) $V=4$, (b) $V=1$, and (c) $V=0.25$.}
\label{fig:ts-V}
\end{figure}

\begin{figure}
\begin{center}
\includegraphics[width=64mm]{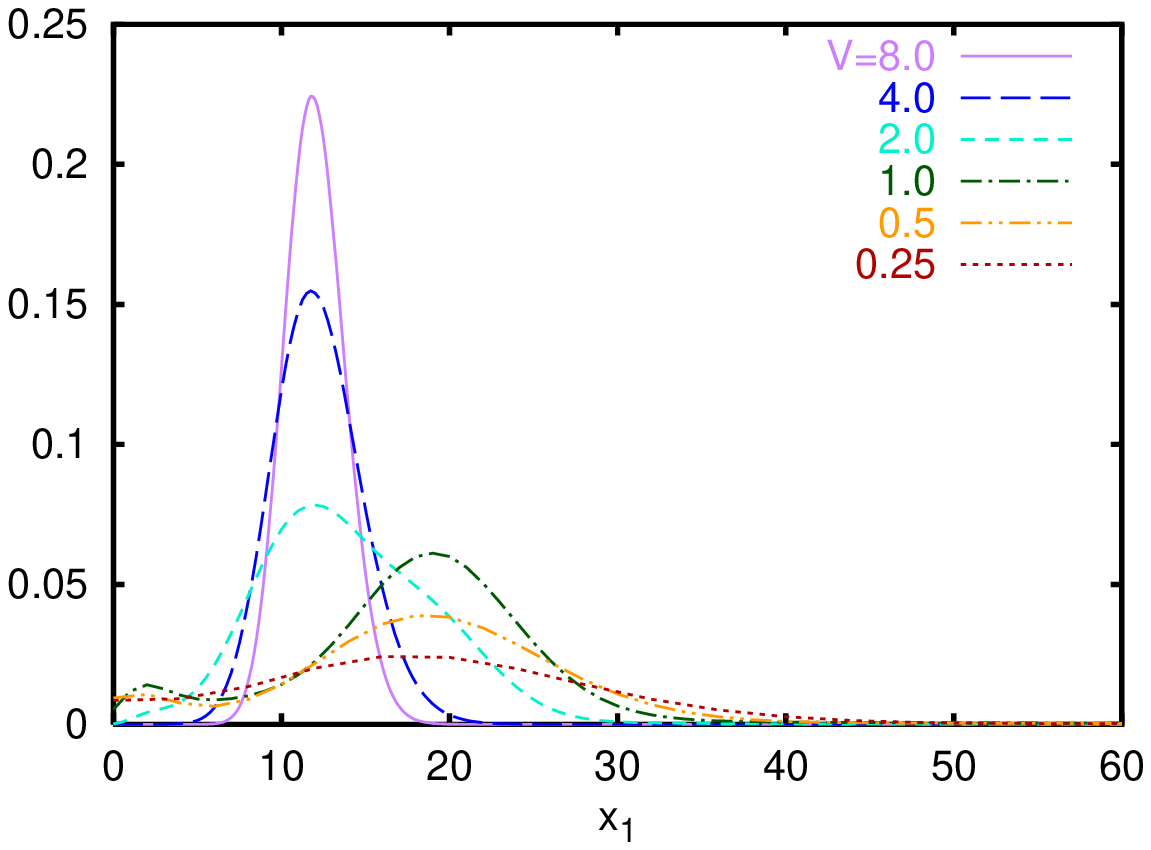}
\includegraphics[width=64mm]{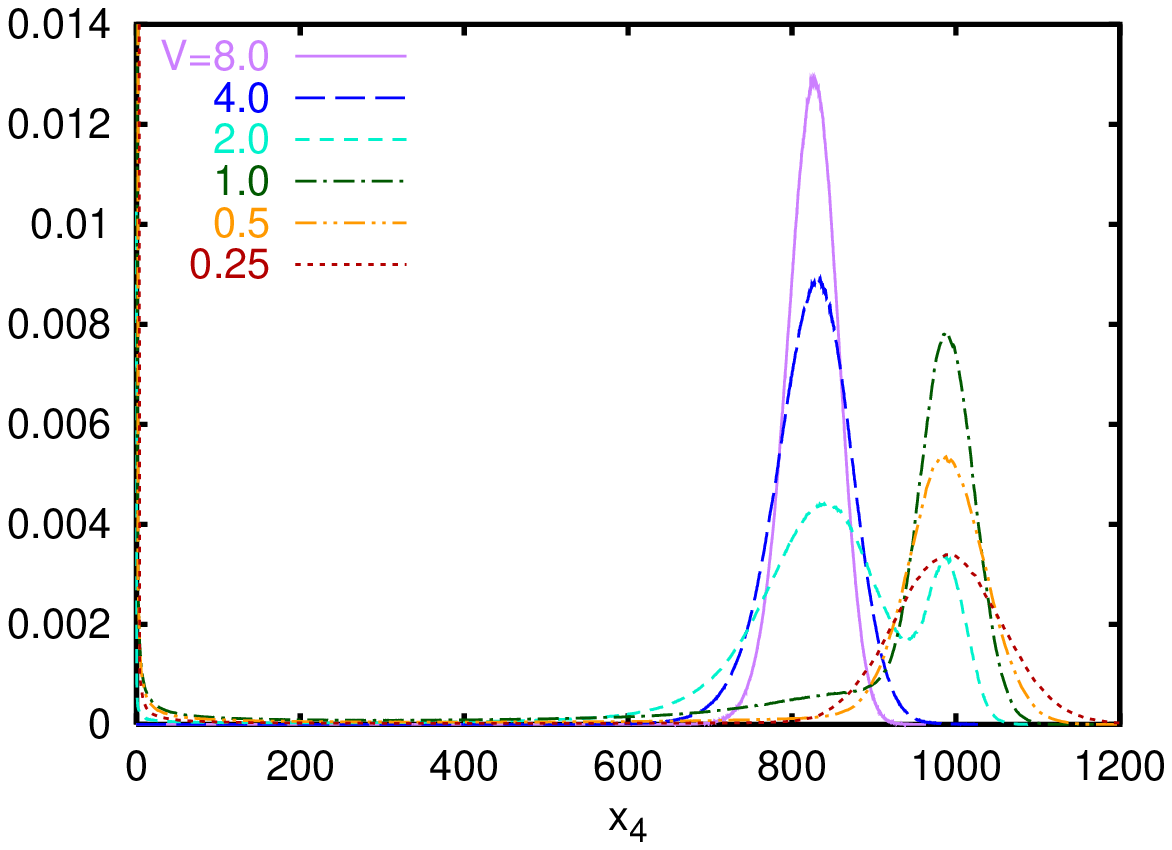}
\end{center}
\caption{Distribution of the concentration $x_{1}$ and $x_{4}$ with
different $V$.
Transition is observed between the $V\le 1.0$ and $V\ge 4.0$ cases.}
\label{fig:xdist}
\end{figure}

Molecular discreteness may not only affect the chemical concentrations
but also the network of reactions.
As seen above, if the number of molecules required for a certain reaction
is $0$, the reaction cannot take place at all.
If such a situation continues for a long time, when compared to the time scales of
other reactions, the system behaves as if the reaction never existed, i.e.,
the reaction is virtually eliminated from the network.
Furthermore, the existence of even one or a few molecules
could cause the resumption of the reaction and the recovery of the network.
In contrast to the continuum limit,
where decay or recovery of the chemical is gradual,
such changes in the network structure are discrete and therefore quick.

Here, we show an example in which the discreteness alters the actual network and
switches the dynamical behavior.
We adopt a simple model with four chemicals and five reactions
among them (see Fig. \ref{fig:net2}) such that
\begin{eqnarray*}
R_{1}:\ X_{1} + X_{4} \stackrel{k_{1}}{\longrightarrow} 2 X_{4} ;&\ &
R_{2}:\ X_{1} + X_{2} \stackrel{k_{2}}{\longrightarrow} 2 X_{2} ;\\
R_{3}:\ X_{2} + X_{3} \stackrel{k_{3}}{\longrightarrow} 2 X_{3} ;&\ &
R_{4}:\ X_{3} + X_{4} \stackrel{k_{4}}{\longrightarrow} 2 X_{4} ;\\
R_{5}:\ X_{4} + X_{2} \stackrel{k_{5}}{\longrightarrow} 2 X_{2} &&
(k_{1}=k_{4}=k_{5}=10^{-3}; k_{2}=k_{3}=10^{-2}).
\end{eqnarray*}
Again, we assume a well-mixed reactor of volume $V$ in contact with a chemical reservoir,
where the concentration of $X_{i}$ is maintained at $s_{i}$ ($D_{i}$ is the flow rate of $X_{i}$).
In the continuum limit, the system is governed by the following rate equations:
\begin{eqnarray*}
\dot{x_{1}} &=& {} - k_{1}x_{1}x_{4} - k_{2}x_{1}x_{2} + D_{1}(s_{1}-x_{1})\\
\dot{x_{2}} &=& \  k_{2}x_{1}x_{2} - k_{3}x_{2}x_{3} + k_{5}x_{4}x_{2} + D_{2}(s_{2}-x_{2})\\
\dot{x_{3}} &=& \  k_{3}x_{2}x_{3} - k_{4}x_{3}x_{4} + D_{3}(s_{3}-x_{3})\\
\dot{x_{4}} &=& \  k_{1}x_{1}x_{4} + k_{4}x_{3}x_{4} - k_{5}x_{4}x_{2} + D_{4}(s_{4}-x_{4})
\end{eqnarray*}
where $x_{i}$ is the concentration of the chemical $X_{i}$.

This reaction network mainly comprises constant flows of chemicals ($R_{1}$ and $R_{2}$)
and an autocatalytic loop ($R_{3}$, $R_{4}$, and $R_{5}$).
Here, we set $D_{i}=D=0.02$ (for all $i$), $s_{1}=10^{3}$, $s_{3}=10$, and $s_{2}=s_{4}=1$.
With these settings, generally,
$X_{1}$ molecules flow into the container and serve as substrates.
They are then converted into other chemicals,
following which they flow out;
this maintains the nonequilibrium condition.
In the continuum limit, the concentrations $x_{i}$ vibrate and converge to the fixed point.

To elucidate the behavior at a condition distant from the continuum limit, we have investigated
the dynamical behavior in such a condition by stochastic simulation.
Fig. \ref{fig:ts-V} shows the time series of $N_{i}$, the number of $X_{i}$ molecules.
When $V$ is large, generally, $N_{i}$ remains large.
This behavior is similar to the rate equation with the addition of noise.
However, when $V$ is small, $N_{i}$ may reach $0$.
In our model, if the system lacks a substrate or a catalyst for a certain reaction,
the reaction ceases completely.
Consequently, the dynamics of the system with such a small $V$ are qualitatively different.

We define the state of the system based on the combination of the reactions that cease.
A system has the following three distinct states (see Fig. \ref{fig:net2}):
\begin{description}
\item[State I.] $N_{i}>0$ for all $i$, and all the reactions occur.\\
This state is determined by the fixed point concentrations obtained by the continuum limit,
and the system converges to the fixed point,
while the vibration around it is sustained
when the number of molecules is finite.
\item[State II.] For the majority of the time,
$N_{2} = 0$, and reactions $R_{2}$, $R_{3}$, and $R_{5}$ cease.\\
The reaction loop cannot proceed,
while reaction $R_{1}$ continuously converts $X_{1}$ into $X_{4}$.
\item[State III.] For the majority of the time,
$N_{4} = 0$, and reaction $R_{1}$ ceases.\\
In the absence of any reactions, the $X_{1}$ molecules accumulate.
An $X_{2}$ molecule flowing in may trigger reactions $R_{2}$ and $R_{3}$ and
convert $X_{1}$ into $X_{3}$.
\end{description}
In the continuum limit, the concentrations cannot reach $0$ due to the constant inflows,
and the system remains at state I when $V$ is sufficiently large,
as shown in Fig. \ref{fig:ts-V} (a),
even though the concentrations fluctuate and vibrate around the fixed point.

With a small $V$, however, the other states appear.
For example, the time series of $N_{i}$ with $V=1$ is shown
in Fig. \ref{fig:ts-V} (b).
At around $t=6500$, the system is in state I,
and it switches to state II at around $t=6700$.
It then alternates between states II and III.
The system spontaneously switches among these states.
If $V$ is considerably smaller, state I is rarely observed,
as shown in Fig. \ref{fig:ts-V} (c).

The distribution of $x_{i}$ is shown in Fig. \ref{fig:xdist}.
A transition is observed with a decrease in $V$.
For a large $V$, the distribution shows a peak at around
$x_{1}=12$ and $x_{4}=8\times 10^{2}$, corresponding to the fixed point of the rate equation.
For a small $V$, the distribution of $x_{4}$ shows peaks at around
$x_{4}=1.0\times10^{3}$ and $x_{4}=0$,
corresponding to state II ($x_{1}\approx 20$, $x_{4}\approx 1.0\times10^{3}$)
and state III (mostly $x_{4}=0$), respectively.

As mentioned above, these states are classified based on
the reactions that cease;
in other words, the states are classified based on
the part of the network that actually functions.
In state I, all the reactions in the network function;
in state II, the autocatalytic loop does not function;
and in state III, the conversion of $X_{1}$ into $X_{4}$ ceases.
The transitions to states II or III can be viewed as
the disconnection of some parts in the reaction network.
Such transitions are possible only if $N_{i}$ reaches $0$,
and therefore, molecular discreteness is essential.
The extinction of the $X_{2}$ and $X_{4}$ molecules makes the system switch to
states II and III, respectively.

The question that arises here is as follows:
In general, which chemicals can switch states in a network?
In our model, molecule $X_{1}$ cannot serve as a switch, even
though $N_{1}$ sometimes reaches $0$ in the case of $V=1$.
First, for a molecule species to function as a key for switching,
$N_{i}$ should be maintained at $0$ for a longer time than that for other reactions.
For $X_{1}$, there is considerable inflow, and the inflow rate is not affected
if $N_{1}$ reaches $0$. Thus, $N_{1}$ cannot remain at $0$ for a long time,
and $X_{1}$ cannot switch the dynamics.
Second, a key chemical for a switch should be located within the reaction paths
and the extinction of the molecule disconnects some reaction paths.

Stochasticity in gene expression is widely studied
with regard to the problem of a small number of molecules in a biological system.
It is often assumed that two states---on and off---are switched by a single regulatory site.
The controlling chemicals and controlled chemicals can be clearly separated.

In contrast, our result shows that chemical species, which are usually
abundant, may sometimes work as a stochastic switch.
In this sense, molecules that are common
or ions such as Ca$^{2+}$ (see \cite{CaOscillator,Thul2004})
may cause stochastic effects.
The role of a chemical may change with time.

\section{Discussion}

In this study, we have demonstrated that molecular discreteness may induce transitions to
novel states in autocatalytic reaction systems, which may result
in an alteration of macroscopic properties
such as the average chemical concentrations.

In biochemical pathways, it is not uncommon to find that the number of
molecules of a chemical is of the order of $10^2$ or less in a cell.
There are thousands of protein species,
and the total number of protein molecules in a cell is not very large.
For example, in signal transduction pathways, some chemicals
work at concentrations of less than $100$ molecules per cell.
There exist only one or a few copies of genetic molecules such as DNA;
furthermore, mRNAs are not present in large numbers.
Thus, regulation mechanisms involving genes are quite stochastic.
Naturally, molecular discreteness involves such rare chemicals.

In the second part of this paper, we have shown that the molecular discreteness may change
the dynamical behavior of reaction networks.
The reaction network is virtually disconnected by the extinction of certain chemicals,
which is not possible in the continuum limit.
Although the network studied here is a small model, similar phenomena can exist
in a complex reaction network with a large number of chemicals and reactions.
We have also investigated random networks of catalytic reactions
$X_{i} + X_{j} \rightarrow X_{i} + X_{k}$ ($j \ne k$).
In such systems, the dynamics of chemical concentrations also depend on the system size $V$.
In a small system, many of the chemical species become extinct ($N_{i}=0$),
and the actual reaction network is disconnected into fragments,
which may be occasionally reconnected by inflow or generation of a molecule.
The onset of change in the concentrations due to disconnection or reconnection is
stochastic and sudden.
In the continuum limit, in contrast, the concentrations gradually converge to
a fixed point in most cases (or to a limit cycle or other attractors).
The simple model in this paper can be viewed as a switching element of such a network;
however, the exact conditions that determine whether a chemical works as a stochastic switch or not
should be addressed in future.

We observed the transitions in the distribution of the concentrations $x_{i}$ with
respect to the change in the system size $V$.
Multiple transitions can also occur, especially if there are
many chemical species for which the number of molecules is sometimes
(but not necessarily always) small.

In this paper, we consider reactions in a well-stirred medium,
where only the number of molecules is taken into account for determining the system behavior.
However, if the system is not mixed well, we need to take into account
the diffusion of molecules in space.
From a biological viewpoint, the diffusion in space is also important
because the diffusion
in cells is not always fast when compared with the time scales of the reactions.
If the reactions are faster than the mixing, we should consider the system
as a reaction-diffusion system, with discrete molecules diffusing in space.
The relation between these time scales will be important,
as indicated by Mikhailov and Hess \cite{Mikhailov1A,Mikhailov2002}.
With regards to these time scales, we recently found that the spatial discreteness
of molecules within the so-called Kuramoto length \cite{Kampen,Kuramoto1},
over which a molecule diffuses in its lifetime
(lapses before it undergoes reaction), can yield novel steady states that are not observed
in the reaction-diffusion equations \cite{YTKK2004,YTKK2005}.
Spatial domain structures due to molecular discreteness are also observed \cite{YTKK2005}.
See also \cite{Shnerb2000,Shnerb2001} for relevance of the discreteness
in a replicating molecule system.

The discreteness-induced effect present here does not depend
on the characteristics of the reactions.
Furthermore, it may be applicable to systems beyond reactions,
such as ecosystems or economic systems.
The inflow of chemicals in a reaction system can be seen
as a model of intrusion or evolution in an ecosystem:
for both systems, discrete agents (molecules or individuals) may become extinct.
In this regard, our result is relevant to the studies of ecosystems,
e.g., extinction dynamics with a replicator model by Tokita and Yasutomi
\cite{Tokita1999,Tokita2003};
strong dependence of the survival probability of new species in evolving networks
on the population size was reported by Ebeling \textit{et al.} \cite{Ebeling}.
The discreteness of agents or operations might also be relevant to
some economic models, e.g., artificial markets.

Most mathematical methods that are applied to reaction systems
cannot appropriately describe the discreteness effect.
Although the utility of simulations with the progress in computer technology
has become convenient,
it would also be important to develop
a theoretical formulation applicable to discrete reaction systems.
On the other hand, in recent years, major advances have been made
in the detection of a small number of molecules
and the fabrication of small reactors,
which raises the possibility to experimentally demonstrate
the discreteness effect predicted here.

We believe that molecular discreteness
has latent but actual importance with respect to biological mechanisms such as
pattern formation, regulation of biochemical pathways, or evolution,
which will be pursued in the future.

\ack
This research is supported by a grant-in-aid for scientific research from
the Ministry of Education, Culture, Sports, Science and Technology of Japan
(11CE2006, 15-11161)
and research fellowships from the Japan Society for the Promotion of
Science (15-11161, abroad H17).


\end{document}